\documentclass{webofc}
\usepackage[varg]{txfonts}   % Web of Conferences font
\newcommand {\ergs}{{\rm erg\ \rm s^{-1}}}
\woctitle{Physics at the Magnetospheric Boundary}
\begin{document}
\title{On the origin of cyclotron lines in the spectra of X-ray pulsars}
%
% subtitle is optionnal
%
%%%\subtitle{Do you have a subtitle?\\ If so, write it here}

\author{A.~A.~Mushtukov\inst{1,2,3}\fnsep\thanks{\email{al.mushtukov@gmail.com}} \and
        J.~Poutanen\inst{1}\fnsep\thanks{\email{juri.poutanen@oulu.fi}} \and
        V.~F. Suleimanov\inst{4,5} \and
        S.~S.~Tsygankov\inst{6,1,7} \and
D.~I.~Nagirner\inst{2} \and
V.~Doroshenko\inst{4} \and
A.~A.~Lutovinov\inst{7} 
        % etc.
}

\institute{Astronomy Division, Department of Physics, PO Box 3000, FI-90014 University of Oulu, Finland
\and
Sobolev Astronomical Institute, Saint Petersburg State University, Saint-Petersburg 198504, Russia 
\and
Pulkovo Observatory of the Russian Academy of Sciences, Saint-Petersburg 196140, Russia
\and
Institut f\"ur Astronomie und Astrophysik, Kepler Center for Astro and Particle Physics, Universit\"at T\"ubingen, Sand 1, 72076 T\"ubingen, Germany
\and
Kazan (Volga region) Federal University, Kremlevskaja str., 18, Kazan 420008, Russia
\and
Finnish Centre for Astronomy with ESO (FINCA), University of Turku,  V\"ais\"al\"antie 20, FI-21500 Piikki\"o, Finland
\and
Space Research Institute of the Russian Academy of Sciences, Profsoyuznaya Str. 84/32, Moscow 117997, Russia
          }

\abstract{%
Cyclotron resonance scattering features are observed in the spectra of some X-ray pulsars and show
significant changes in the line energy with the pulsar luminosity. 
In a case of bright sources, the line centroid energy is anti-correlated with the luminosity. 
Such a behaviour is often associated with the onset and growth of the  
accretion column, which is believed to be the origin of the observed emission and the cyclotron lines. 
However, this scenario inevitably implies large gradient of the magnetic field strength 
within the line-forming region, and it makes the formation of the observed line-like features problematic. 
Moreover, the observed variation of the cyclotron line energy is
much smaller than could be anticipated for the corresponding luminosity changes.
We argue that a more physically realistic situation is that the cyclotron
line forms when the radiation emitted by the accretion column is reflected from
the neutron star surface. The idea is based on the facts that a substantial part of column luminosity is
intercepted by the neutron star surface and the reflected radiation should contain absorption features. 
The reflection model is developed and applied 
to explain the observed variations of the cyclotron line energy  
in a bright X-ray pulsar V~0332+53 over a wide range of luminosities.
}
\maketitle
\section{Introduction}
\label{intro}
X-ray pulsars are neutron stars in binary systems accreting matter usually from a massive companion. 
A strong magnetic field channels accreting gas towards magnetic poles
and modifies the observed X-ray spectrum 
manifesting as the line-like absorption  features, the so-called cyclotron lines. 
Such cyclotron resonance scattering features (CRSF), 
sometimes also with harmonics, are observed 
in the spectra of  several X-ray pulsars \cite{Coburn2002,Fil2005,CW12}. 

In some cases, the luminosity related changes of the line energy are observed,
suggesting that  configuration of the line-forming region depends on the
accretion rate. The line energy has been reported to be positively-correlated with luminosity in
relatively low-luminosity sources \cite{Staub2007,Kloch2012} 
and negatively-correlated in high-luminosity sources \cite{Mih1998,TsL2006}, 
as well as uncorrelated with it \cite{CP13}.  

Here we will focus only on
the high-luminosity case. The negative correlation of the CRSF energy with
luminosity is usually explained with the onset and growth of the 
accretion column at high luminosities \citep{BS1976}. In this scenario, the height
of the column, and, therefore, the average displacement of the emission and the
line-forming regions from the   neutron star surface increase with luminosity,
which to a shift of CRSF to lower energies. 
The problem is, however, that the predicted shift is much larger than the observed one. 
The column height depends on luminosity almost linearly \citep{BS1976} and the magnetic field weakens with distance 
as $r^{-3}$, and yet brightening by more than an order of magnitude yields at most 25\% decrease
in the CRSF energy \citep{TsL2006,TsLS2010}. 
Moreover, large gradient of the magnetic field and of the accretion velocity are 
expected to smear out the line-like features making it difficult to explain why we observe CRSFs at all.
  
\begin{figure*}
\center{\includegraphics[width=0.95\linewidth]{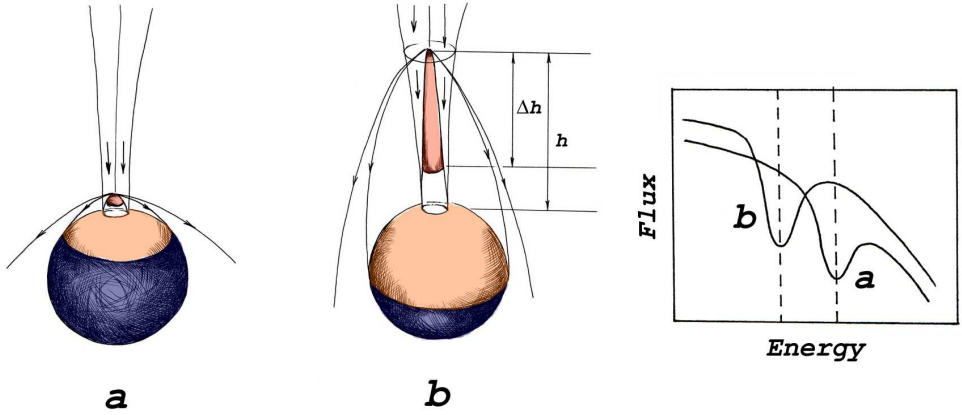}}
\caption{Accreting X-ray pulsar geometry and the emergent spectrum.  
The larger is the luminosity, the higher is the accretion column, the larger illuminated fraction 
of the neutron star surface, the weaker the average magnetic field, 
and the smaller the  cyclotron line energy.}
\label{fig:sketch}
\end{figure*}

We propose another scenario \cite{P2013}. We show that a significant part of the radiation from 
the accretion column should be intercepted by the stellar 
surface because of the relativistic beaming (Section \ref{sec:refl} and \citep{KFT76,LS1988}),
and the absorption features should exist in the radiation reflected from the 
magnetised atmosphere (Section \ref{sec:line}).
Therefore, it is natural to assume that  the line is 
formed in the atmosphere of the neutron star due to reflection of the intercepted radiation. 
In this case the negative correlation between the luminosity and the cyclotron line energy is reproduced
because of the changes in the illuminated part of a stellar surface: if the $B$-field decreases away from the magnetic poles,
the relatively high column could illuminate regions 
with relatively low field strength (see Fig.~\ref{fig:sketch}).

\section{Model set up and reasons to the model}
\label{sec:model}

The physical picture of the accretion on the magnetised neutron star is discussed in classical 
papers \cite{BS1976,KFT76}.  
At low accretion rate, free-falling material heat the neutron star 
surface near its magnetic poles, and the bright spots radiate energy in the X-ray range. 
At high accretion rate, the radiation pressure stops the infalling material
above the neutron star surface in a radiation-dominated shock, after which 
the gas slowly sinks down towards the surface radiating the dissipated energy 
through the side walls of the accretion column.
The column is expected to arise when the
luminosity exceeds a critical value \citep{BS1976}:
\begin{equation} \label{eq:limLum}
L^{*}\approx 4\times10^{36}\left(\frac{\kappa_{\rm T}}{\kappa_{\parallel}}\right)
\left(\frac{5l}{R}\right)
\left(\frac{M}{M_{\odot}}\right)\ \ergs,
\end{equation}
where $\kappa_{\parallel}$ is the electron scattering opacity along the magnetic field,
$\kappa_{\rm T}$ is the Thomson opacity, $M$ and $R$ are the mass and the radius of the star, 
$l$ is the length of the accretion arc at the stellar surface.

The column height  depends on the mass accretion rate almost linearly \citep{BS1976,LS1988}:
\begin {equation} \label{eq:ColHeigh}
\frac{h}{R}=\dot{m}\ \ln\left(\eta\frac{1+\dot{m}}{\dot{m}^{5/4}}\right),\quad
\eta=\left(\frac{B^2 d^2\kappa_{\parallel}}{7\pi c\sqrt{2GMR}}\right)^{1/4}  
= 16 \left(\frac{B}{5\times 10^{12}\ \mbox{G}} \right)^{1/2} \left(\frac{d}{100\ \mbox{m}}\right)^{1/2} 
\left(\frac{\kappa_{\parallel}}{0.4} \right)^{1/4},
\end{equation}
where $d$ is the thickness of the accretion arc, and  
$\dot{m}=L/L^{**}$ is a ratio of the X-ray pulsar luminosity to the limiting 
luminosity for the magnetised neutron star  
\begin {equation} \label{eq:EddLum}
L^{**}\approx  10^{39}
\left(\frac{l/d}{50}\right)
\left(\frac{\kappa_{\rm T}}{\kappa_{\parallel}}\right)
\left(\frac{M}{M_{\odot}}\right)\ \ergs .
\end{equation}  
Thus, the luminosity of X-ray pulsars in a bright state is in the range $[L^{*},L^{**}]$.

\subsection{Reflected fraction}
\label{sec:refl}

Since the radiation energy density drops off sharply towards the edge of the column, the height, 
where matter stops, varies inside the accretion channel and depends on the distance
from its edge.
As a result, the height is maximal in the middle of the channel and decreases 
towards the borders.
Therefore, the
radiation from the already stopped matter should pass through a layer of the rapidly falling plasma, 
which is not supported by the radiation and falls down with velocity close to the 
free-fall velocity $\beta=v/c=\sqrt{r_{\rm S}/r}$ (here  $r_{\rm S}=2GM/c^2$ is the  Schwarzschild radius). 
The optical thickness of these layers is high enough to change dramatically the angular distribution of the 
emergent radiation, and due to the relativistic beaming it is directed mainly towards the stellar surface.  
For the case of electron-scattering dominated column, the angular distribution of the column luminosity in the 
laboratory frame is given by \cite{P2013}:
\begin {equation} \label{eq:LSd}
\frac{dL (\alpha) }{d\cos\alpha} = 
 I_0 \ \frac{D^4}{\gamma} 2 \sin\alpha  
\left(  1 +\frac{\pi}{2} D \sin\alpha\right),
\end{equation}
where $\alpha$ is the angle between the photon momentum and the velocity vector,  
$\gamma=1/\sqrt{1-\beta^2}$ is the Lorentz factor,  
$D=1/[\gamma(1-\beta\cos\alpha)]$ is the Doppler factor, and $I_0$ is the normalization constant.

Thus, radiation from the accretion column is captured partly by the stellar surface, and its fraction
 $L_{\rm c}/L$ could be substantial because of the beaming.
The fraction obviously depends on the height of the column, brightness distribution over the column, 
radiation beam pattern and the compactness of the star.
Fraction of the captured radiation for a case of point source above the neutron star surface
as a function of the height-to-radius ratio is given in Fig.~\ref{fig:Part} and has quite high 
values even in a case when the height is comparable with the neutron star radius.

\subsection{Formation of the cyclotron line in reflected radiation}
\label{sec:line}

Radiation from the accretion column reaches the neutron star surface and is reflected. 
The most important process affecting the spectrum of the reflected radiation is Compton scattering. 
The cross-section for Compton scattering in a strong magnetic field is 
energy-dependent and has strong resonances. 
Photons at the resonance energies cannot penetrate deep into the atmosphere, they interact in the surface 
layers and scatter back changing the energy. This produces the lack of the photons in the line core.
In the cyclotron line wings, photons penetrate deeper into the atmosphere and scatter there also changing
their energy. If they scatter into the resonance energy, they cannot leave the atmosphere because of 
the larger optical depth there and escape instead in the line wings. 
Thus, the lack of the photons near the resonance is not filled in and manifests itself as an 
absorption features in the spectrum.
The absorption features at the harmonic energies can be even stronger, because the absorbed 
photons are mostly reemitted at the energy of the fundamental. 
The typical X-ray pulsar spectrum cuts off at $\sim$30--50 keV, and therefore 
the contribution of the harmonics will be negligible (as is in the case of V~0332+53). 

%Thus, estimation of the reflected fraction and the existence of the absorption feature in the reflected radiation
%give us a proof that the line forms on the stellar surface after reflection of the column radiation from the 
%atmosphere of a neutron star.

\section{The line energy and comparison with the observations}

\subsection{The line position}

It is possible to estimate a line centroid energy in the context of the reflection model. 
Neglecting the asymmetry in the line shape, the  
centroid of the CRSF in the  reflected spectrum averaged over the surface and all angles is
determined by the $B$-field strength weighted with the
distribution of flux, $F(\theta)$, 
and the  line equivalent width over the neutron star surface. 
Assuming that the line equivalent width in the reflected radiation is constant over  the surface, 
the cyclotron line centroid energy is proportional to the mean field 
\begin{equation} \label{eq:E_ave}
E_{\rm cycl} \propto 
\langle B\rangle=\frac{\int\limits_{0}^{\pi}B(\theta)F(\theta)\sin\theta\ d\theta}
{\int\limits_{0}^{\pi}F(\theta)\sin\theta\ d \theta}.
\end{equation} 

Note that the model immediately gives a limitation for changes in the cyclotron line energy.
Because for the dipole field, the $B$-field strength drops only by a factor of two from the pole to the equator,
the line should lie in the energy interval $[E_0/2;E_0]$, where $E_0$ is the energy corresponding to the polar field $B_0$. 
A more realistic lower limit on the line energy can be obtained assuming a  uniformly illuminated surface, i.e. $F(\theta)={\rm const}$. 
In that case
\begin {equation}
\langle B\rangle_{\min}=\frac{1}{4}B_0\int\limits_{0}^{\pi}\sin\theta\sqrt{1+3\cos^2\theta}\ d\theta\simeq0.7 B_0,
\end{equation} 
and, therefore, the model predicts the range of $[0.7E_0;E_0]$ for 
the line centroids is in agreement with observations \citep{TsL2006,TsLS2010}. 
This prediction provides a test for the model.

Thus, the problem of estimation of a line energy comes to the calculating of the flux distribution over the
stellar surface. The standard deviation of the magnetic field  $\sigma_B$ gives an estimate of the minimum width of the line, since
variations of the magnetic field over the surface lead to its smearing.
In our calculations we approximate the geometry of the accretion column by a thin stick at the magnetic pole. 
This approximation is reasonably accurate for high $B$-field pulsars up to $h\lesssim R$.
We also assume that most of the energy is emitted in a region of characteristic scale $\Delta h$ 
situated above the neutron star surface at height $h$ (see Fig.~\ref{fig:sketch}).

\begin{figure}[h]
\begin{center}
\begin{minipage}[h]{0.45\linewidth}
\includegraphics[width=1\linewidth]{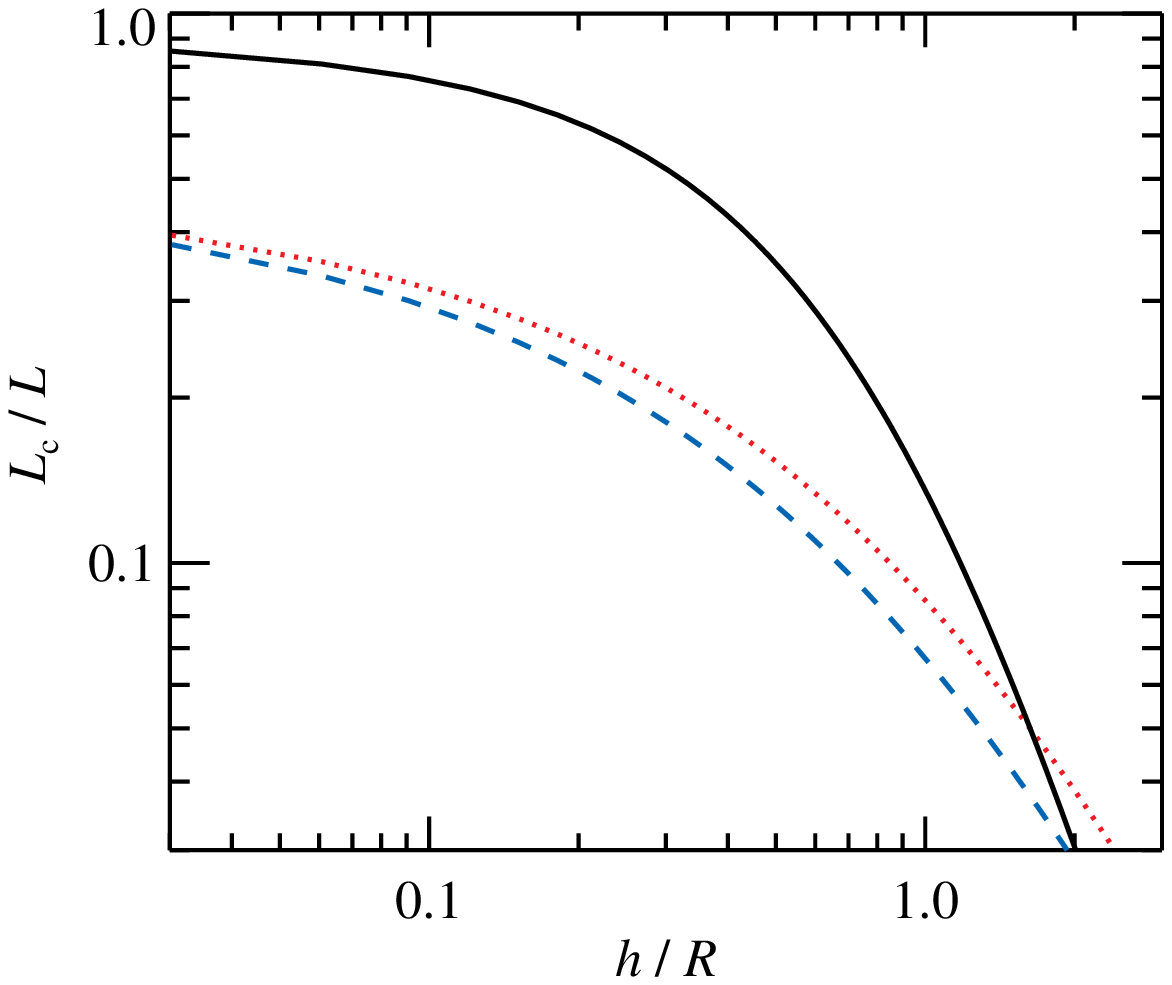}
\caption{Fraction of the captured radiation from a point source above the neutron star surface
as a function of the height-to-radius ratio.
The dashed blue curve and the dotted red curve correspond to the isotropic source in 
flat space-time and in  Schwarzschild metric correspondingly.
The solid black  curve corresponds to the beaming pattern given by equation (\ref{eq:LSd}).  
Here $R=3 r_{\rm S}$.} 
\label{fig:Part}
\end{minipage}
\hfill 
\begin{minipage}[h]{0.45\linewidth}
\includegraphics[width=1\linewidth]{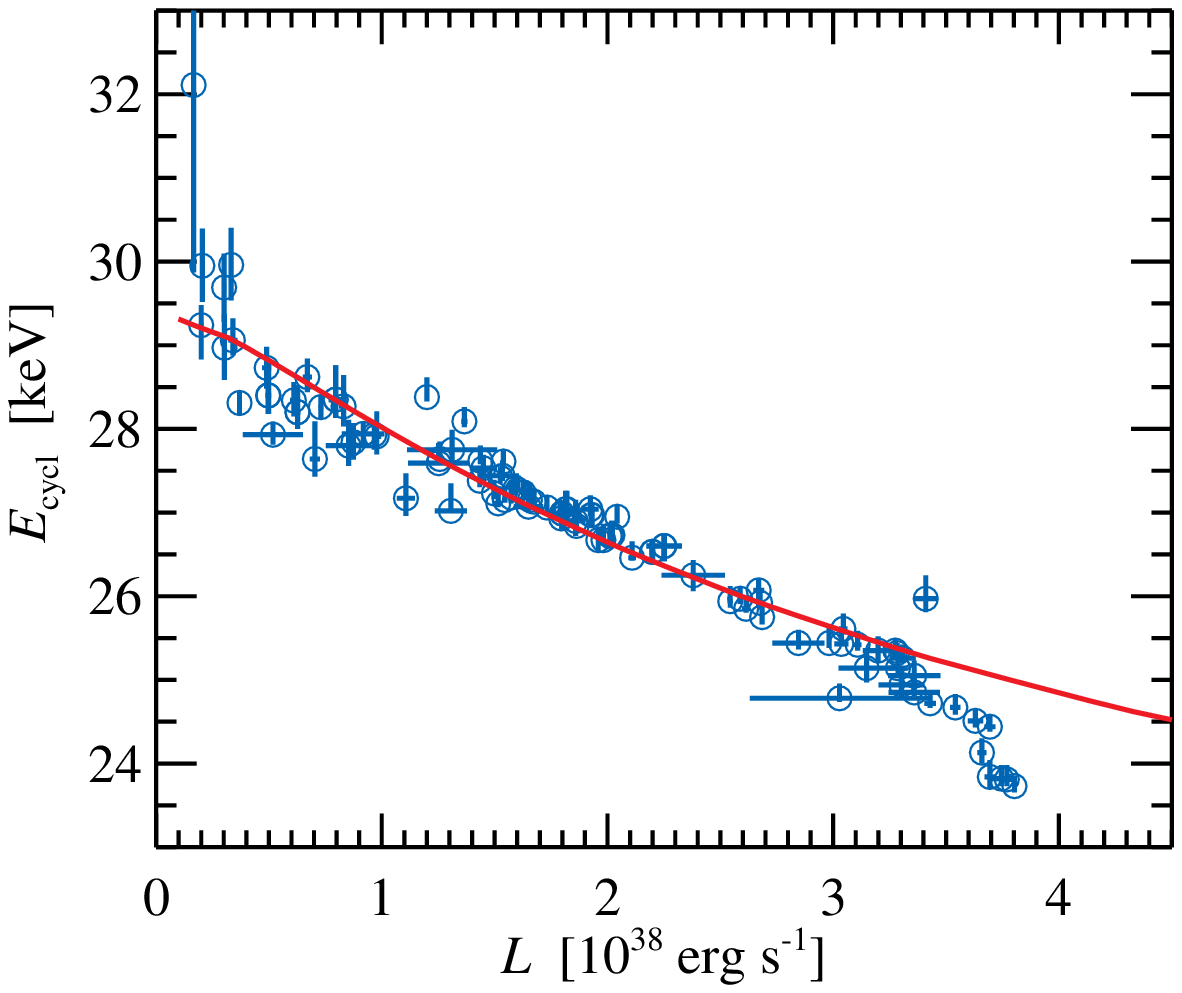}
\caption{Dependence of the cyclotron line energy on the luminosity in 
the X-ray pulsar V~0332+53 (circles with the error bars; from \cite{TsLS2010}) and 
the best-fit theoretical relation obtained with the model (solid line). The fit parameters
are: $\Delta h/h=0.1$,  
$\eta=15$, $E_0=29.5$~keV, and $L^{**}=21\times 10^{38}\ \ergs$.
\smallskip\smallskip\smallskip
}
\label{fig:col_LS}
\end{minipage}
\end{center}
\end{figure}

\subsection{Comparison with  observations}
\label{sec:obs}

The model was compared with the data from the X-ray pulsar V~0332+53 obtained 
with the \textit{RXTE} and \textit{INTEGRAL} observatories during outburst in 2004--2005 
\citep{TsL2006,TsLS2010}. The data set has 
the information about the behavior of this object in a wide range of the luminosities,
from $\sim10^{37}$ up to $\sim4\times10^{38}\ \ergs$ and shows negative correlation between
the luminosity and the centroid of the CRSF.
 
The best-fit theoretical relation of the cyclotron-line energy dependence on the luminosity for 
the X-ray pulsar V~0332+53 \citep{TsLS2010} is shown in Fig.~\ref{fig:col_LS}. 
It is clear that the data at luminosities above $3.5\times 10^{38}\ \ergs$ cannot 
be described by the model and we neglect them in the fits. 
The remaining data are fitted with two free parameters $E_0$ and $L^{**}$. 
We fix $\eta=15$, because the results depend on that parameter very weakly
and vary $\Delta h/h$ in the range between 0 and 1.
The values of the limiting luminosity, $L^{**}$, obtained in the fits should be taken as the upper limits 
on the actual value given by equation (\ref{eq:EddLum}): 
radiation from the upper part of the column might be completely blocked by the 
falling material and photons escape at a height significantly smaller than the top of the 
accretion shock because of their drift in the falling plasma. 

The fitting procedure shows that the best description of the data is achieved for a column 
with most of the emission coming from its top. This does not necessarily mean that the lower part is 
not emitting, but rather that this emission does not hit the neutron star surface. 
Furthermore, our results are based on the assumption of the constant with latitude equivalent width of a line. 
Thus, our result of the top-dominated emission from the column might 
be an artefact of this effect.

%%%%%%%%%%%%%%%%%%%%%%%%%%%%%%%%%%%%%%%%%%%%%%%%%%%%%%%%%%%%%%%%%%%%%%%%%%%%%%
%% DISCUSSION AND/OR CONCLUSIONS                                            %%
%%%%%%%%%%%%%%%%%%%%%%%%%%%%%%%%%%%%%%%%%%%%%%%%%%%%%%%%%%%%%%%%%%%%%%%%%%%%%%
 
\section{Summary}
\label{sec:summary}

A reflection model for the cyclotron line formation in the spectra of X-ray pulsars is proposed. 
It is based on two facts: a large part of the column radiation is intercepted 
by the neutron star surface and the absorption feature forms in the reflected radiation.  
Small variations of the magnetic field along the surface imply that the line 
centroid energy can vary by at most 30\% for a dipole field, 
and it makes model easily testable (i.e. it can be ruled out if larger variations are observed).   
Changes in the pulsar luminosity are expected to
anti-correlate with the average magnetic field over the illuminated surface and, therefore, 
with the line centroid energy, exactly as observed during the outburst of the X-ray pulsar V~0332+53.
Our model explains well the observational data and has profound  implications for the interpretation 
of the data on the cyclotron lines observed in X-ray pulsars. 

The model implies that the cyclotron line should exhibit variations of its energy and equivalent width with
the pulsar phase, since the observer would see different parts of the illuminated surface during a pulse period. 
The specific model predictions are in the agreement with the pulse-resolved observational data
obtained for the X-ray pulsar V~0332+53 \cite{L2013}.

In order to predict the line parameters more accurately, a detailed model of the reflection of the 
column radiation from the atmosphere is required. It will provide a possibility of detailed diagnostic
of X-ray pulsars in a bright state.

%\bibliographystyle{woc}
%\bibliography{proc}

\end{document}